\documentclass[aps,prd,amsmath,longbibliography]{revtex4-1}

\usepackage[dvips]{graphicx}
\usepackage{latexsym,epsfig,bm,psfrag,subfigure}
\usepackage{color}         
\usepackage[bookmarksnumbered,bookmarksopen,colorlinks,citecolor=blue,linkcolor=blue]{hyperref} %

\newcommand{\eq}[1]{Eq.~(\ref{#1})}

\newcommand{\bfnabla}{\mbox{\boldmath$\nabla$}}
\def\bea{\begin{eqnarray}}
\def\eea{\end{eqnarray}}
\def\g{\gamma}\def\t{\tau}\def\s{\sigma}\def\o{\omega}
\def\bfp{{\bf  p}}
\def\la{\langle}\def\ra{\rangle}\def\d{\delta}\def\k{\kappa}\def\z{\zeta}
\def\o{\omega}\def\bfb{{\bf b}}
\def\bfv{{\bf v}}\def\r{\rho}
\def\bfx{{\bf x}}
\def\bfP{{\bf P}}\def\wt{{2\k^2\over P^+}t}
\begin{document}



\title{NT@UW-21-04\\Color Transparency and  the Proton Form Factor- Feynman Wins}


\author{Olivia  Caplow-Munro   }
\email{petryolivia@gmail.com}

\author{Gerald A. Miller}
\email{miller@uw.edu}
\affiliation{Department of Physics, University of Washington, Seattle, WA \ \ 98195, USA}

\date{\today}

\begin{abstract}
A recent experiment  [{\bf Phys. Rev. Lett. 126,082301 (2021)}] used the $(e,e'p)$ reaction on $^{12}$C to search for the effects of color transparency (the absence of final state interactions).
Color transparency was said to be ruled out. The ability to observe the effects of color transparency depends on the ability of a putative point-like-configuration (PLC), formed in a high-momentum transfer coherent reaction, to escape the nucleus without expanding its size.
 We study the expansion aspect of color transparency  using   superconformal baryon-meson symmetry and light-front holographic QCD. A new formalism is obtained and used to analyze the recent experiment.    
The resulting conclusion  is that effects of  expansion  would not be sufficiently significant in causing final state interactions to occur. Therefore we conclude that a  PLC was not formed. This means  that the  Feynman mechanism  involving virtual photon absorption on a single high momentum quark is responsible for high momentum electromagnetic form factor of the proton.
 \end{abstract}




\maketitle
\section{Introduction}
The recent striking  experimental finding~\cite{Bhetuwal:2020jes}, that color transparency does not occur in reactions with momentum transfer  up to  14.2 GeV$^2$ demands an interpretation and assessment of the consequences. This paper is aimed at providing such, beginning with a brief discussion of color transparency. \\

Strong interactions are strong: when hadrons hit nuclei they generally break up the nucleus or themselves. Indeed, a well-known semi-classical formula states that the intensity of a beam of hadrons falls exponentially with the penetration distance through nuclei.  This effect is known as absorption.
It is remarkable that  QCD
admits the possibility that, under certain specific conditions, the strong interactions can effectively be turned off and hadronic systems can move freely through a nuclear medium.\\

How can this occur? Consider that elastic scattering of a small color-singlet $q\bar q $ system by a nucleon. At sufficiently high energy, the separation $b$ of the quarks (in a direction transverse to the momentum) does not change. The lowest-order perturbative contribution is given by two-gluon exchange \cite{Low:1975sv,Nussinov:1975mw,Gunion:1976iy} and the  remarkable feature is that, in the limit that $b$ approaches 0, the cross section vanishes because color-singlet point particles do not exchange colored gluons. This feature is expressed concisely as $\lim_{b\to0}\s(b^2)\propto b^2.$  This reduced interaction, caused by interference between emission by quarks of different colors in coherent processes, is the basic ingredient behind QCD factorization proofs and is widely used \cite{Donnachie:2002en} and not questioned.\\
 
It is natural  to suppose that small-sized color-singlet objects stand out in processes involving coherent high momentum transfer reactions. Early perturbative QCD calculations~\cite{Farrar:1979aw,Efremov:1979qk,Lepage:1979za,Lepage:1979zb,PhysRevD.21.1636,DUNCAN1980159,PhysRevD.21.1636},
 see Fig.~1a, of the pion elastic electromagnetic form factor were interpreted \cite{Mueller:1982bq} in the following manner:
A high-momentum, $Q$, photon hits one of the partons that  greatly increases the relative momentum to $Q$. The system can only stay together only by exchanging a gluon carrying that momentum. That gluon has a range of only  $1/Q$ so that the partons must be close together, making a small-sized system or point-like configuration (PLC).
As time goes by the point-like configuration  evolves to a   physical wave function of the pion. Many  consequences of the PLC idea were spelled out by Frankfurt \& Strikman~\cite{Frankfurt:1988nt}. A clear analysis of the proton in perturbative QCD  was presented in~\cite{Lepage:1980fj}. \\

 \begin{figure}[h] \label{Diags}

		\includegraphics[width=.54\textwidth]{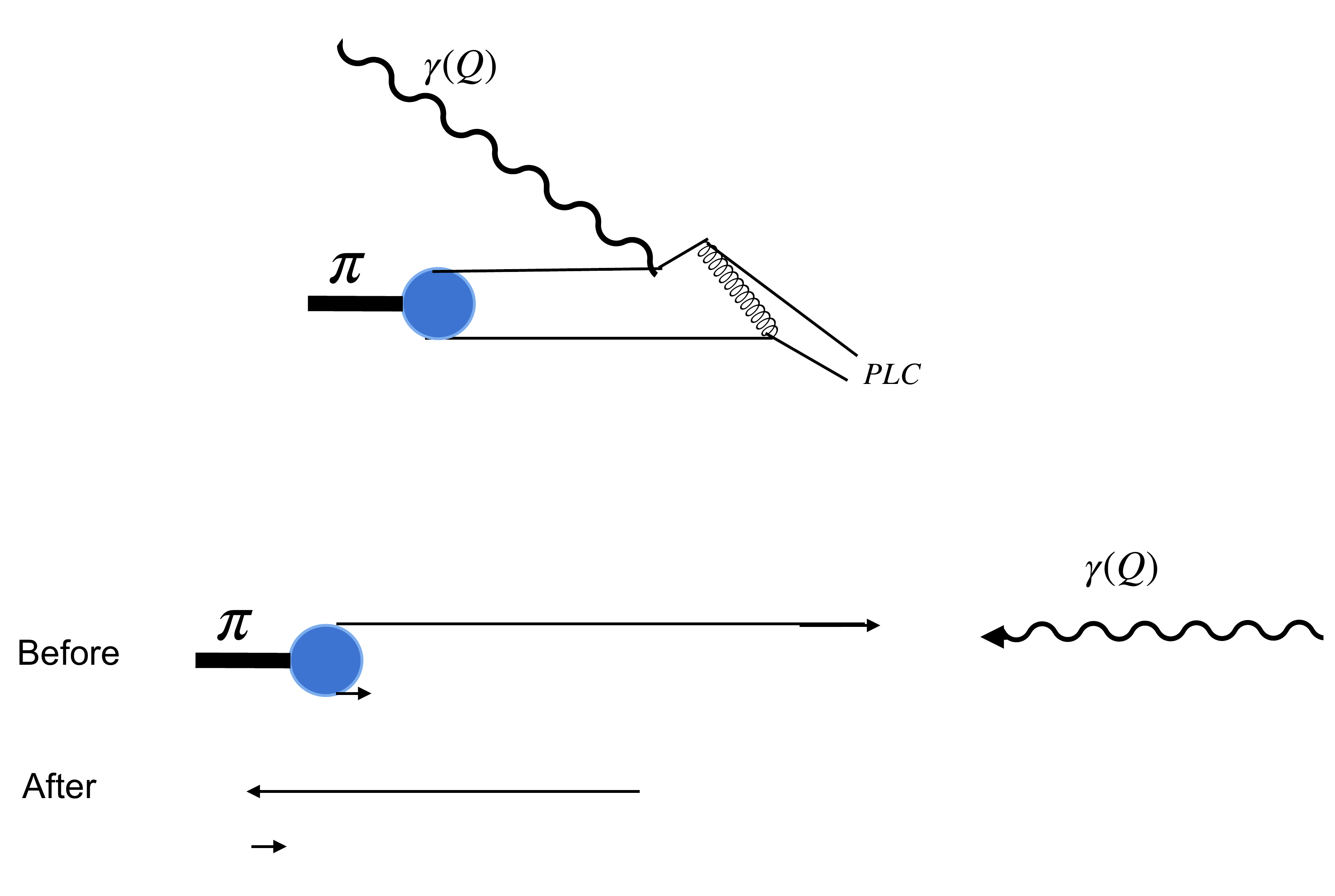} 
     \caption{ High momentum transfer reaction mechanisms. Top picture: a pQCD mechanism. Other diagrams of the same order are not shown. Middle picture: Initial state in the Feynman mechanism. Bottom pictire: final state in the Feynman mechanism. The final state has a good overlap with the turned around version of the initial state.}
      \end{figure}  
A particular  consequence is that if a high-momentum transfer reaction  that is coherent on  the hadronic level  occurs in a nucleus then the initial and final state reactions that normally occur are suppressed. 
This is the phenomenon of color transparency. However, it is important that the putative PLC not expand to a system of ordinary hadronic size when it is in the nucleus.  The expansion time in the rest frame of the ejected proton is of the order of the inverse of an excitation energy. The effects of time dilation increase this time as the proton momentum  increases. For this reason, the  very  high-momentum transfer experiment~\cite{Bhetuwal:2020jes} is of  vital importance.\\

Although it is natural to suppose that PLCs dominate  coherent high-momentum transfer
processes, it is far from obvious that this is the case~\cite{PhysRevD.21.1636,DUNCAN1980159,PhysRevD.21.1636,Feynman:1973xc}.  The doubts are especially relevant for baryons. Indeed, Feynman \cite{Feynman:1973xc} remarked that ``if a system is made of 3 particles, the large $Q^2$ behavior depends not  on the singularity when just two come together, but rather when all three are on top of one another". Furthermore, ``such pictures are too simple and inadequate".\\

Another   mechanism competes with  PLC dominance  (Fig.~1b). In this idea, first published by Drell \& Yan \cite{Drell:1969km} (but known as the Feynman mechanism \cite{Feynman:1973xc}) the process occurs when a single quark, carrying a large fraction of the nucleon's momentum, is turned around by the  incident virtual photon. The spectator system is not required to shrink to a small size and color transparency effects involving protons would not be expected to occur. In this picture, a reasonably valid  relation between elastic and deep inelastic scattering is obtained. See also~\cite{West:1970av}. A more recent example is ~\cite{PhysRevD.58.114008} that favors the Feynman mechanism.\\

 Models  of Generalized Parton Distributions (GPDS)~\cite{ Ji:2004gf} access both the longitudinal and transverse structure of nucleons so that measurements thereof can distinguish the different mechanisms. It has been said~\cite{Diehl:2003ny} that GPDs parameterize soft dynamics akin to the Feynman mechanism.   Specific models of GPDs, for example~\cite{deTeramond:2018ecg} also favor the Feynman mechanism.  More generally, a review of the  history  teaches us ~\cite{Belitsky:2005qn} that there are only two proposals for the mechanism responsible for  high momentum elastic reactions. \\

To summarize~\cite{Jennings:1990ma}, the magical disappearance of a strong scattering amplitude arises in  three steps: 
(a) A coherent high-momentum transfer reaction creates  a color-singlet PLC  (b)  the PLC doesn't interact (a well-defined aspect of QCD) 
(c) If the PLC escapes the nucleus before expanding, then no re-interactions occur and effects of color transparency are to be observed. \\

The non-interaction of a PLC is a given, as is the necessity of expansion. The truly interesting aspect of color transparency is whether or not PLC's are created in high-momentum transfer coherent reactions. However, evaluating the effects of expansion creates significant uncertainty in making conclusions. This paper is devoted to providing a better estimate of expansion that is based on new findings that use holographic  techniques to represent the effects of confinement on excited hadronic states, as discussed in the review~\cite{Brodsky:2014yha}.\\

The considerable efforts to observe color transparency, along with a particular success~\cite{Aitala:2000hc},  has been reviewed several 
times~\cite{Frankfurt:1994hf,Jain:1995dd,Ashery:2006zw,Dutta:2012ii}. Here we  focus on color transparency studies involving
incident high-energy  pions at FermiLab~\cite{Aitala:2000hc}, electro-production of pions at JLab~\cite{Clasie:2007aa}, electro-production of rho mesons at JLab~\cite{ElFassi:2012nr}, and efforts in the $(e,e'p)$ reaction that culminated with the recent JLab experiment~\cite{Bhetuwal:2020jes}. The FermiLab experiment involved pions of $500 $ GeV. The $(e,e'\pi)$ experiment ~\cite{Clasie:2007aa}
involved space-like squares of virtual photon four-momentum $Q^2$ between 1.10 and 4.69. GeV$^2$, with outgoing  pions of corresponding momentum between 2.79 and 4.4 GeV. The $(e,e'\r)$ experiment involved  $Q^2$ between 1 and 3 GeV$^2$, with virtual photon energy around 3 GeV, so as to avoid the effect of changing the coherence length. The outgoing momentum of the $\r$-meson was about 5 GeV for all kinematic conditions. \\

The signature of color transparency in the FermiLab experiment was a spectacular dependence of the cross section for di-jet production by incident pions on the nuclear atomic number of about $A^{5/3}$, for high relative transverse momentum between the two jets,  in accord with theoretical predictions~\cite{Frankfurt:1993it,PhysRevD.65.094015}. The electron scattering experiments measure the transparency, $T$, which is the ratio of the measured experimental yield to the expected yield in the absence of final interactions. The signature of the onset of color transparency is a rise in the value of $T$ with respect to an increase with $Q^2$. A significant rise can occur only if the effects of color transparency are present.\\

The aim here is to evaluate the effects of expansion by using relativistic light-front wave functions obtained from
light front holographic QCD, an approach explained in the review~\cite{Brodsky:2014yha},  providing a relativistic treatment of confined systems.  Light-front quantization is a relativistic, frame-independent approach   to describing  the constituent structure of hadrons. The simple structure of the light-front (LF) vacuum allows an unambiguous definition of the partonic content of a hadron in QCD and of hadronic light-front  theory~\cite{Brodsky:1997de}.  The spectrum and light-front wave functions of relativistic bound states are obtained in principle from the eigenvalue equation  $H_{LF} \vert  \psi \rangle  = M^2 \vert  \psi \rangle$ that becomes an infinite set of coupled integral equations for the LF components.  
 This provides a quantum-mechanical probabilistic interpretation of the structure of hadronic states in terms of their constituents at the same light-front time  $x^+ = x^0 + x^3$, the time marked by the front of a light wave~\cite{Dirac:1949cp}.      The matrix diagonalization~\cite{Brodsky:1997de} of the frame-independent  LF Hamiltonian eigenvalue equation in four-dimensional space-time has not  been achieved, so    other methods and approximations ~\cite{Brodsky:2014yha}  are needed to understand the nature of relativistic bound states in the strong-coupling regime of QCD.\\

To a first semiclassical approximation, where quantum loops and quark masses are not included, the relativistic bound-state equation for  light hadrons can be reduced to an effective LF Schr\"odinger equation. The technique is to  identify the invariant mass of the constituents  as a key dynamical variable, 
 $\zeta$, which measures the separation of the partons within the hadron at equal light-front  time~\cite{deTeramond:2008ht}.  Thus,  the multi-parton problem  in QCD is reduced, in   a first semi-classical approximation, to an effective one dimensional quantum field theory by properly identifying the key dynamical variable.  In this approach the complexities of the strong interaction dynamics are hidden in an effective potential $U$. \\

It is remarkable that in the semiclassical approximation described above,  the light-front Hamiltonian  has a structure which matches exactly the eigenvalue equations in AdS space~\cite{Brodsky:2014yha}.  This offers the  possibility to explicitly connect  the AdS wave function $\Phi(z)$ to the internal constituent structure of hadrons. In fact, one can obtain the AdS wave equations by starting from the semiclassical approximation to light-front QCD in physical space-time.   This connection yields a  relation between the coordinate  $z$ of AdS space with the impact LF variable $\zeta$~\cite{deTeramond:2008ht},  thus giving  the holographic variable $z$ a precise definition and intuitive meaning in light-front QCD.\\

Light-front  holographic methods  were originally introduced~\cite{Brodsky:2006uqa,Brodsky:2007hb} by matching the  electromagnetic current matrix elements in AdS space~\cite{Polchinski:2002jw} with the corresponding expression  derived from light-front  quantization in physical space-time~\cite{Drell:1969km,West:1970av}. It was also shown that one obtains  identical holographic mapping using the matrix elements of the energy-momentum tensor~\cite{Brodsky:2008pf} by perturbing the AdS metric  around its static solution~\cite{Abidin:2008ku}, thus establishing a precise relation between wave functions in AdS space and the light-front wave functions describing the internal structure of hadrons.\\

The light front wave functions that arise out of this light front holographic approach provide a new way to study old problems that require the  use of relativistic-confining quark models. The study of  the expansion of a PLC is an excellent example of such a problem. \\

The starting point for the present calculations is the relativistic light-front wave equation for hadronic light front wave  function $\phi(\z)$ that encodes the dynamical properties.  Consider a  hadronic bound state of  two constituents, each of vanishing mass.   
The   wave equation for a hadronic bound state of two constituents  is given by: 
	\bea 
	\bigg(-{d^2\over d\z^2}-{1-4L^2\over 4\z^2}+U(\z,J)\bigg)\phi(\z)=M^2\phi(\z)
	\label{we} \eea	
	The  eigenmodes of this equation determine the hadronic mass spectrum and represent the probability amplitude to find the partons at transverse impact separation $\z$. The connection between $\phi$ and wave functions $\Phi$ in a higher-dimensional space asymptotic to anti-de-Sitter space is explained in Ref.~\cite{Brodsky:2014yha}.
In the   soft-wall model  ~\cite{Karch:2006pv}	the potential is harmonic and holographically related to a unique dilaton profile the follows from the requirements of conformal invariance. The effective potential is given~\cite{Brodsky:2014yha} by 
 	\bea U(\zeta,J) = \kappa^4\zeta^2+2\k^2(J-1)\label{sw}\eea
	where $ \kappa $ is the strength of the confinement,  and $J$ is the total angular momentum. 
	This  form was used in ~\cite{PhysRevLett.102.081601,PhysRevD.91.045040,PhysRevD.91.085016}, in particular  to obtain relations between the baryon and meson spectrum~\cite{PhysRevD.91.085016}, with a universal value of $\k=0.53$ GeV. The results are specified here to: $U(\z)=\k^4 \z^2 +C_H$, with $C_\pi=-2\k^2,\,C_\r=0,\,C_N=2\k^2$. We shall see that these different values have profound effects on the ability to observe color transparency.
	The associated eigenvalues are :
	$M^2_{nL} = 4\kappa^2(n+{J+L_M\over 2}) $	for mesons, and for baryons
$ M^2_{nL} = 4\kappa^2(n+L_B+1) $
with $L_M=L_B	+1$.\\

The next step is to convert \eq{sw} to a light front 3-dimensional wave equation in coordinate space.
This is done by using
$ \phi(\z)\to \sqrt{\z}\psi(x,\bfb)= \sqrt{z}e^{iL\phi}\psi_L(x,b)$ with the replacement $\z\to b \sqrt{x(1-x)}$,
so that the equation for a meson is given by:
\bea \bigg(-{1\over x(1-x) }\nabla_b^2+U(b^2(x(1-x),J)\bigg)\psi(x,\bfb)=M^2 \psi(x,\bfb)\label{lfe}
\eea
The equation for the nucleon is of the same form, but there are two components, denoted $\psi_{\pm}$. The smaller-sized of these two components
is $\psi_+$. Therefore  that is the relevant one whose time development is studied here. A good description of electromagnetic form factors and parton distributions is obtained~\cite{Brodsky:2016yod}.\\

Note that the Hamiltonian contains   no operator that changes the value of $x$. The longitudinal  degree of  freedom is fixed in this model. However, the known meson and baryon spectra are  well  reproduced, which is the central necessity for obtaining a reasonable treatment of the expansion of the PLC wave packet.
The  light-front wave functions obtained  by  	solving \eq{lfe} 
are given by:
	\bea \psi_{nL_z}(\bfb,x) =  {\sqrt{2} }\sqrt{n!\over (n+L)!} {e^{i L_z \varphi} \over\sqrt{2\pi}}e^{-\k^2b^2 x(1-x)/2}(\sqrt{x(1-x)}b)^{L} L_n^{L}(\k^2 b^2 x(1-x)),\label{wf}\eea
		where  $ b $ is the transverse separation, $ \varphi $ is the transverse angle, and $ x $ is the plus-momentum ratio of one of the quarks.\\  
%

The average value of $b^2$ in the ground state is given by  $\bar b^2\equiv\la \Psi_{00}|b^2|\Psi_{00}\ra={1\over \k^2 x(1-x)}.$ This quantity can be very large if $x$ approaches 0 or 1. However, there would be an $x$-dependent wave function that vanishes at those endpoints in a more complete treatment. Here we are concerned with studying how a small-sized wave packet, or PLC expands as a function of time. This means that cases with $x$ near 0 or 1 are not under consideration. Moreover, if $x$ is near the endpoints, the Feynman mechanism  is the dominant contributor to the  form factor, so that the conclusion of this paper would be reached without the need to study expansion.\\

It is necessary to obtain the time-development operator for a system that is moving in the $z$ direction with a given large value of $P^+$. 
This will be done using the Feynman path integral technique~\cite{Feynman:100771}.
The first step is to regard the right-hand-side  of \eq{sw} as an operator $P^-P^+$, to obtain the LF Schroedinger equation
\bea 2i{\partial\over \partial \t}\Psi=P^- \Psi={1\over P^+} \bigg(-{1\over x(1-x) }\bfnabla_b^2+U(b^2(x(1-x),J)\bigg)\ \Psi ,\label{lfh}\eea
with the light-front time variable: $\tau\equiv x^+$.
The factor of 2 arises from  the definition: $ x^\pm=x^0\pm x^3$, so that $P\cdot x={1/2}(P^-x^++P^+x^-) -\bfP\cdot\bfx$. The vectors written in bold face are two-dimensional vectors often written with the subscript $\perp$.\\

  The time development of a PLC that moves at high speed is studied here.
The longitudinal momentum fraction, $x$, is taken as fixed in the present model. This is a good approximation because interactions proportional to $b^2$ do not  change the value of $x$~\cite{Donnachie:2002en,PhysRevLett.68.17}.\\

 One could construct the time-development operator, $K$, using the stated wave functions of \eq{wf}. Instead, it is much more efficient to use the Feynman path integral formalism~\cite{Feynman:100771}, in which $U$ is the path integral of the  exponential of $i$ times  the classical action.
 In this formalism the  Lagrangian is expressed in terms of velocities that are defined as partial derivatives of the Hamiltonian with respect to momenta. The entire formalism is based on  a Hamiltonian operator that is $i$ times the partial derivative with respect to time. For this purpose, it is useful to define a time $t\equiv  \t/2$. For a system moving  in the laboratory with speed  close to that of light, $x^3\approx x^0$, so that $t$ is approximately the laboratory time. \\

The next step is to identify a Lagrangian that corresponds to the above Hamiltonian. The operator $\bfp= -i \bfnabla_b $
so that by Hamilton's equations  ${\partial\over \partial t}\bfb= \bfv={\partial H\over \partial \bfp}=2\bfp/(P^+x(1-x))$. The following expression results:
\bea L={P^+\over4}x(1-x)\bfv^2-{1\over P^+} U(\zeta,J)\eea
The formalism can be tested by taking the limit of a free particle, $U(\zeta,J)\to 0$, using the well-known expression for the time development of a plane wave, starting at $t=0$.  At later times, the initial plane wave acquires  time-dependence of  
$e^{-i P^-t}= e^{-i P^-\tau/2}$, which is the correct form.\\

Having specified the
Lagrangian,   the known result  \cite{Feynman:100771}
 for  harmonic oscillator potentials lets us  obtain the time development operator $K$ for times  between 0 and $t$:
\bea 
 \la \bfb,t|K|\bfb',0\ra=\bigg[{x(1-x)\k^2\over 2\pi i \sin {2\k^2t\over P^+}}\bigg]\exp\bigg[i\big({x(1-x)\k^2\over 2 \sin{2\k^2t\over P^+}}\big)((b^2+{b'}^2)\cos {2\k^2t\over P^+}-2\bfb\cdot\bfb')\bigg]\exp{(-i{C_H \over P^+}t)}.\label{U}\eea
  		The term $C_H$ accounts for the constant term in $U$, and $C_\pi=-2\k^2,\,C_\r=0,C_N=2\k^2$.
One may check that the above expression becomes $\d(\bfb-\bfb') $ in the limit that $t\to0$.  The same limit is achieved if $P^+\to\infty$ in which 
the separation between partons in the PLC is unchanged as it moves through the nucleus. This fixed transverse  separation configuration is said to be frozen.    Note that the presence of the term $C_H$ shows that the expansion process depends on the hadron that is involved. This non-universality of the
expansion process  that arises from a unified treatment of meson and baryon spectra is  the first new result presented here.  See also
Ref.~\cite{FARRAR1990125}.\\

Given the time-development operator  of \eq{U}, we may  study the time development of a PLC, potentially formed in a high momentum transfer (hard) reaction. A measure of expansion was introduced in Ref.~\cite{Frankfurt:1993es}. The idea is that a PLC is originated via a  hard interaction involving nucleons initially bound in a nucleus.
The  soft interactions between the PLC and the surrounding medium  is proportional to the square of the transverse separation distance \cite{Low:1975sv,Nussinov:1975mw,Gunion:1976iy}.
We wish to compare the 
 relative effects 
 of escaping with an interaction to that of escaping without  an interaction. This is given by  a ratio defined as $b^2(t)$.
 The effective size is given by the ratio of matrix elements:
\bea b^2(t)\equiv {\la \Psi_{00}| b^2 K(t)|\rm PLC\ra\over \la \Psi_{00}|\rm PLC\ra},\label{bdef}\eea where $\Psi_{00}$ is the ground state wave function of \eq{wf}, and $K(t)$ is the time-development operator of \eq{U}. This expression is first-order in a final-state interaction.  A  complete multiple scattering series was presented in~\cite{PhysRevD.47.1865}. The term $i\s {b^2(t)\over \la b^2 \ra}$ has been thought of as scattering amplitude that varies along the path length, $\ell$, of an outgoing PLC, with $t=\ell$~\cite{Frankfurt:1988nt}.\\

Evaluations  of $b^2(t)$  proceed by taking a simple form for the PLC: $\la \bfb|\rm PLC\ra=e^{-\g b^2}$. Evaluation   of \eq{bdef} is then straightforward because  Gaussians are involved. 
 The result is that 
\bea b_H^2(t)={2\over\k^2} {i\over  x(1-x)}\sin{(\wt)}\exp{[-i({4\k^2\over P^+}+{C_H\over P^+})t]} +{2\over\k^2x(1-x)+2\g}\exp{[-i({6\k^2\over P^+}+{C_H\over P^+})t}].\label{res}\eea
The time dependence is  seen to be independent of the value of $x$, and  the results are periodic with an angular frequency of 
$\o\equiv 2\k^2/P^+$.  As a result, the second term vanishes smoothly for all values of $t$ as $\g$ approaches infinity, the point-like limit.
In that case, the first term of \eq{res} accurately describes the time development and 
 is independent of the  detailed form of  $\la \bfb|\rm PLC\ra$, provided the size parameter of the PLC is sufficiently small.

The expression for $b^2_H(t)$ is complex-valued (as has been known for a long time~\cite{Jennings:1990ma}), but only the real part contributes to 
the absorption of the outgoing ejected proton. Therefore, we take the limit $\g\to\infty$ and use the 
  real part of $b^2_H$. This and using  the stated values of $C_H$  yields  the results:
\bea&{b^2_\pi\over 2\bar b^2}= \sin^2(\wt) \\
&{b^2_\r\over 2 \bar b^2}= \sin(\wt) \sin({4\k^2\over P^+}t)\\
&{b^2_N\over 2\bar b^2}= \sin(\wt)\sin({6\k^2\over P^+}t). 
\eea
These expressions contain the remarkable result that the effects of confinement cause the expansion time to be  very different for pions, rho mesons and nucleons.   The oscillations inherent in the above results are of no consequence. Once the first maximum is hit the system is strongly absorbed and the formulae are no  longer relevant.\\

 \begin{figure}[h] \label{pion}

		\includegraphics[width=.4\textwidth]{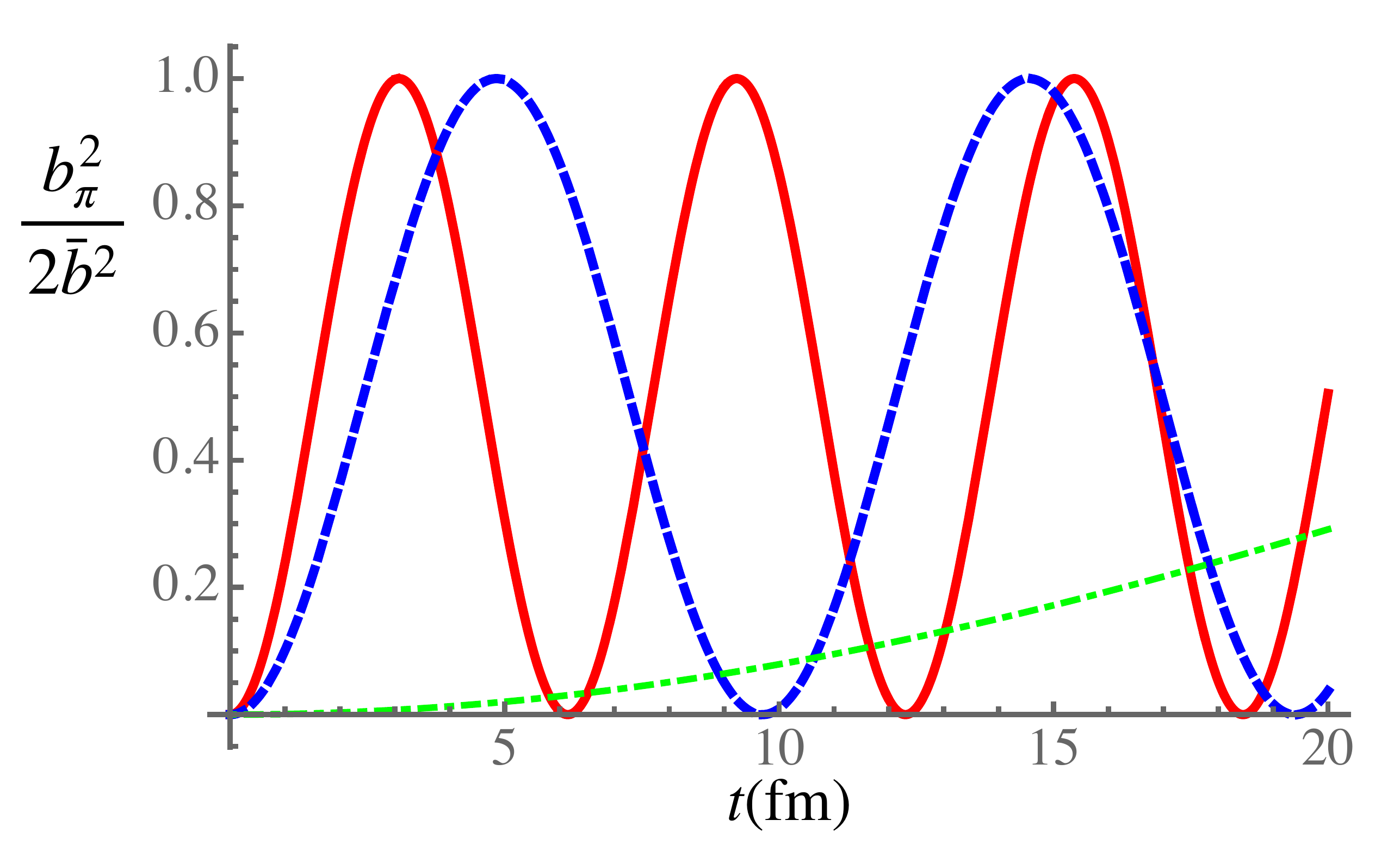} 
     \caption{${b^2_\pi\over 2\bar b^2}$. Solid (red)  $P_\pi^+=5.5 $ GeV, Dashed (blue)   $P_\pi^+=8.8 $ GeV, Dot-dashed (green) $P_\pi^+=100 $ GeV. $t$ is in units of fm}   
  \label{rho}
		\includegraphics[width=.4\textwidth]{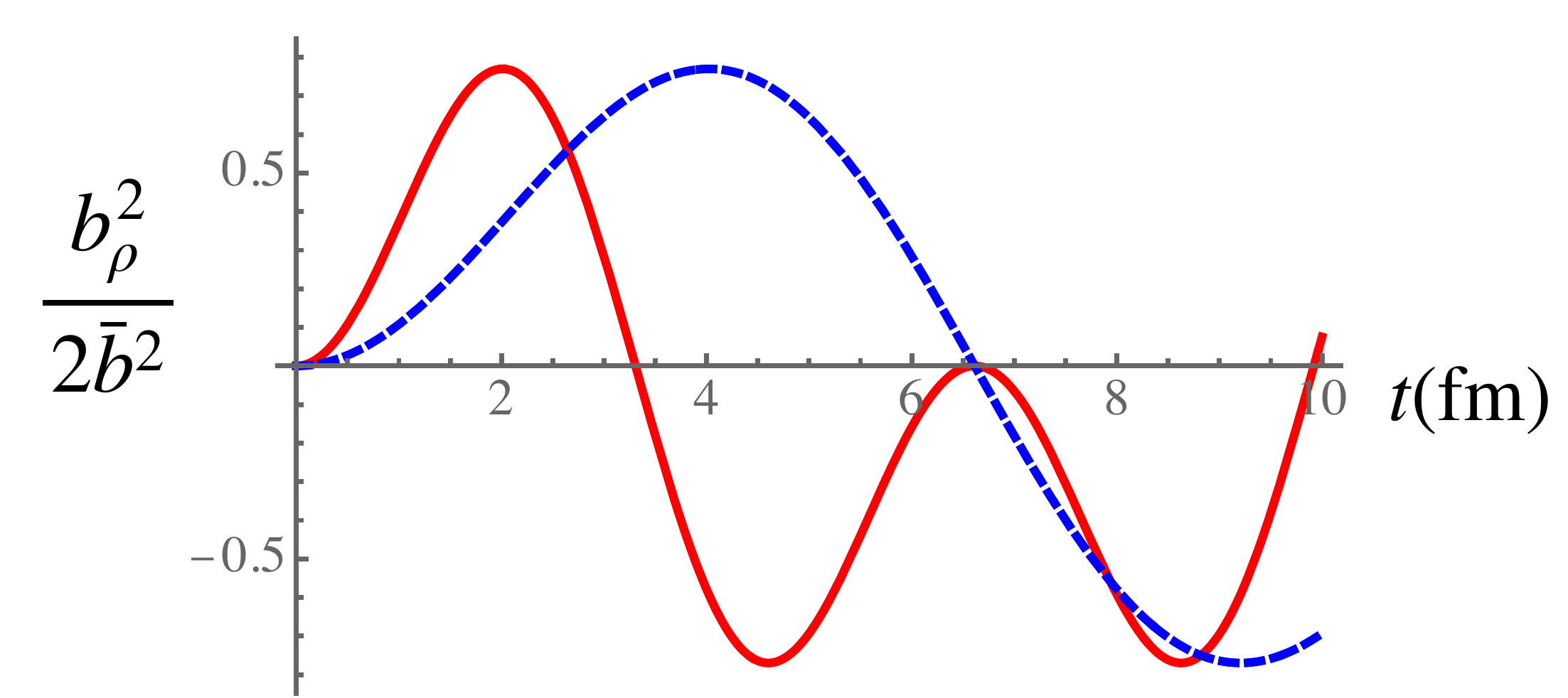} 
     \caption{${b_\r^2\over 2\bar b^2}$. Solid (red)  $P_\r^+=6 $ GeV, Dashed (blue)     $P_\r^+=12 $ GeV. $t$ is in units of fm}\includegraphics[width=.4
    \textwidth]{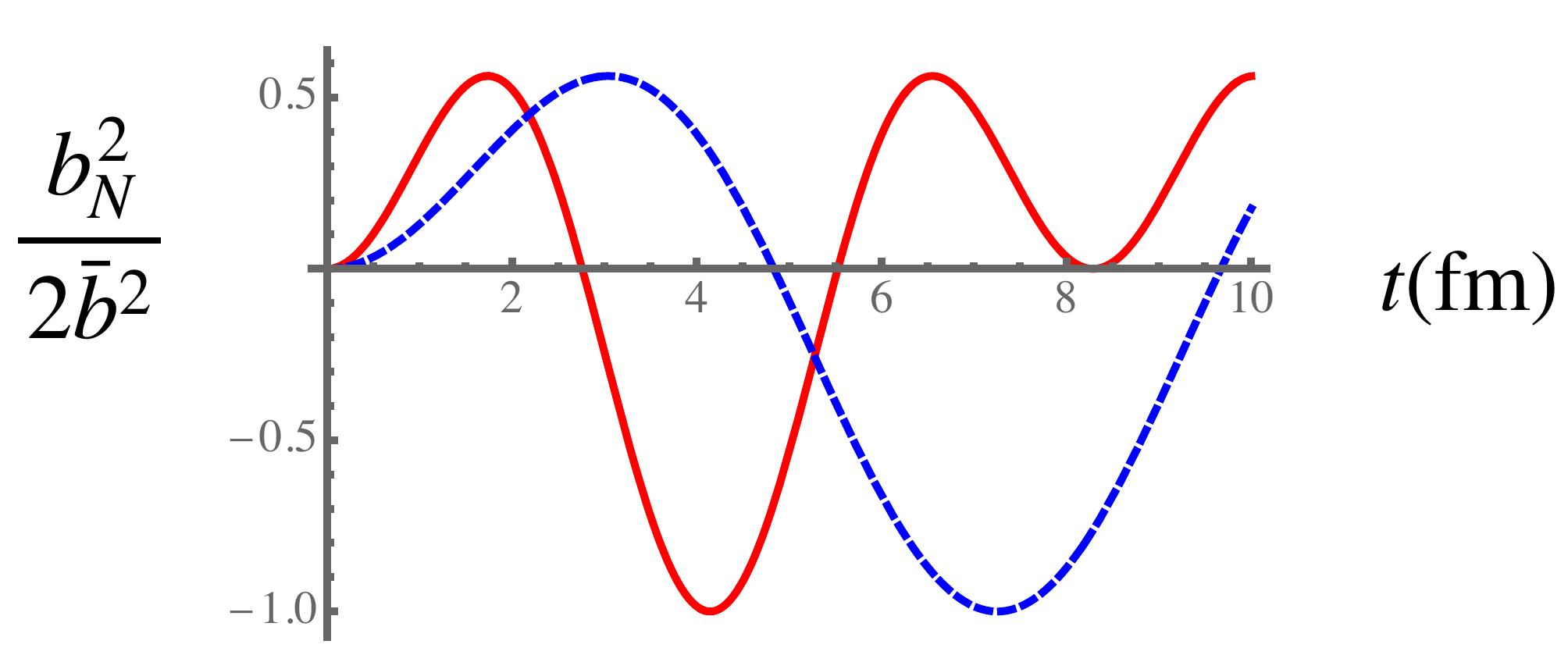} 
     \caption{${b_N^2\over 2\bar b^2}$  Solid (red) $P_N^+=8 $ GeV,   Dashed (blue)   $P_N+=14 $ GeV. $t$ is in units of fm. }  \end{figure}  
The results of evaluations are shown in Figs.~2-4. 
 The pionic results of Fig~2 are presented for the upper and lower values of $P_\pi^+$ of Ref. \cite{Clasie:2007aa} along with 100 GeV/c. It is useful to define the expansion time, $t_H$ as the  {\it lowest}  time $t_E$ for which $b^2_H$ hits its maximum value.  For the pion this is
$t_\pi = 1.57{P^+\over 2\k^2} =0.552 {P^+\over \rm GeV} \rm fm.$ This time increases from about 2 fm to 6 fm in the kinematic range of Ref.~\cite{Clasie:2007aa} . This increase is large enough so that the observed increase in the measured transparency can reasonably be ascribed to the effects of color transparency. The FermiLab experiment was done at 500 GeV, but the results shown here are for 100 GeV. This is done only to allow a non-zero value to be shown. It is very clear that, in the FermiLab experiment, the PLC goes through an entire nucleus without expanding.\\


The results for the expansion of a  $\r$ PLC are presented in Fig.~3. For the experiment of~\cite{ElFassi:2012nr} $t_\r= .955{P^+\over 2\k^2}=0.335{P^+\over \rm GeV} \rm fm$ is only about 2 fm for all of its kinematic variables. This indicates that the observed rise in the transparency, while consistent with theoretical predictions, may not be due to color transparency. However, a new experiment that doubles the value of $P^+$ could lead to observation of the effects of color transparency.\\

The key results here are for the proton as shown in Fig.~4.
The proton expansion time is given by  $t_p = 0.659{P^+\over 2\k^2} =0.231{P^+\over \rm GeV} \rm fm.$ In the $(e,e'p)$  experiment on a $^{12}$C target~\cite{Bhetuwal:2020jes}, the value of $Q^2$ ranges between 8 and 14.2 GeV$^2$. The photon energy $\nu= Q^2/(2M)$, and $P^+_p\approx 2\nu$. Thus $P^+$ ranges from about 8 to 14 GeV. As shown in Fig.~3, $t_p$ ranges between about 2 to about 3 fm. This is large enough to observe the effects of color transparency because the  radius of $^{12}$C  is only about 2.4 fm.  If a PLC had been formed, it would have made it out of the nucleus without being absorbed. Thus we are led to the inevitable conclusion that a PLC was not formed. This means that all of the quarks are   not on top of each other during the high-momentum transfer process, so that the Feynman mechanism is the only one left standing.\\

 The conclusion that the Feynman mechanism is responsible for the proton electromagnetic form factor raises a number of interesting questions regarding its implications. The first is whether this Feynman dominance occurs for all hadrons, in particular the pion. The pion electromagnetic form factor has been studied extensively via perturbative and non-perturbative means, including lattice QCD calculations.  In the analytic examples, both the perturbative QCD and light front holography calculations lead to an asymptotic form factor $\sim 1/Q^2$. The PLC is part of the former calculations. The Feynman mechanism does dominate in the latter calculations at infinite momentum~\cite{Sheckler:2020fbt}. However,      values of $x$ near 1/2   dominate the light front wave function~\cite{Brodsky:2007hb}  for the  relative momenta involved in   the FermiLab experiment. Thus the result that the astounding $A$-dependence observed is due to color transparency is not challenged.\\
 
 Another point concerns the relation with the EMC effect, the nuclear suppression of high quark structure function at large values of Bjorken $x$. Frankfurt \& Strikman suggested~\cite{Frankfurt:1985cv,Frankfurt:1988nt} that this suppression is due to the reduction of the nucleon's  PLC component caused by the attractive nature between the nucleon's non-PLC component and the residual nucleus. This idea has been followed up in many  papers, including~\cite{Frank:1995pv,CiofidegliAtti:2007ork,Hen:2016kwk,Miller:2020eyc}, and an excellent interpretation along with  a qualitative description of nuclear structure functions in the valence region has been obtained. The question arises: does the lack of PLC dominance of the electromagnetic form factor cast doubt on the basic idea that the different nuclear interactions of different-sized components of the nucleon wave function is responsible for the EMC  effect. The answer is no, because the having a  dominant PLC is not necessary. The only necessity is that different sized components of the nucleon wave function interact differently with the nucleus.  A first application of this idea that uses    a  light front holographic treatment~\cite{deTeramond:2018ecg} of the nucleon wave function was presented in ~\cite{Miller:2020eyc}. The two different components are configurations of three and four partons. Further development is proceeding.\\
 
The results of Figs.~2-4 arise from using a specific model, based on a semi-classical approximation,  in which  quantum loops and quark masses are not included. For example, situations in which the starting configuration includes gluonic components are not included. In that case,  the high-momentum transfer process proceeds via a mechanism that is not the Feynman mechanism. This would correspond to a new, unpublished  mechanism that is neither perturbative QCD nor the Feynman mechanism.\\
 
The  present results provide a new way to compute the effects of the spatial expansion of a putative PLC. The  formalism of the light front holographic model and the resulting expansion effects, combined with the experimental result~\cite{Bhetuwal:2020jes} that color transparency does not occur in reactions with momentum transfer  up to  14.2 GeV$^2$, leads to an answer to a decades old question: the Feynman mechanism dominates the high-momentum transfer values of the proton electromagnetic form factor.

 

 \section*{Acknowledgements}
This work was supported by the U. S. Department of Energy Office of Science, Office of Nuclear Physics under Award Number DE-FG02-97ER- 41014. We thank Mark Strikman for a very useful discussion.
%

\end{document}